\documentclass{article}
\usepackage{graphicx}  
\usepackage{amsmath}   
\usepackage{amssymb}   
\usepackage{bm} 
\usepackage{dcolumn}
\usepackage{color}
\usepackage{mathrsfs}
\usepackage{amsfonts}
\usepackage{varioref}
\RequirePackage[colorlinks,citecolor=blue,urlcolor=magenta,linkcolor=blue]{hyperref}
\labelformat{section}{Section #1} 
\labelformat{subsection}{Section #1} 
\labelformat{subsubsection}{Section #1}
\labelformat{subsubsubsection}{Section #1}
\labelformat{equation}{Eq.~(#1)} 
\labelformat{figure}{Fig.~#1} 
\labelformat{subfigure}{Fig.~\thefigure#1} 
\labelformat{table}{Tab.~#1} 
\labelformat{appendix}{Appendix #1}

\title{\bf Particle creation by de Sitter black holes revisited}
\author{Sourav Bhattacharya\footnote{sbhatta[AT]iitrpr.ac.in}\\
{{ Department of Physics, Indian Institute of Technology Ropar}}\\ {{Rupnagar, Punjab 140 001,
India}}}
\begin{document}
 
\maketitle
\begin{abstract}
Creation of thermal distribution of particles by a black hole  is independent of the detail of gravitational collapse, making the construction of the eternal horizons suffice to address the problem in asymptotically flat spacetimes. For eternal de Sitter black holes however, earlier studies have shown the existence of both thermal and non-thermal particle creation, originating from the non-trivial causal  structure of these spacetimes. Keeping this in mind we consider this problem in the context of a quasistationary gravitational collapse occurring in a $(3+1)$-dimensional eternal de Sitter, settling down to a Schwarzschild- or Kerr-de Sitter spacetime and consider a massless minimally coupled scalar field.   There is a unique choice of physically meaningful `in' vacuum here, defined with respect to the positive frequency cosmological Kruskal modes localised on the past cosmological horizon ${\cal C^-}$, at the onset of the collapse.  We define our `out' vacuum at a fixed radial coordinate `close' to the future cosmological horizon, ${\cal C^+}$, with respect to positive frequency outgoing modes written in terms of the ordinary retarded null coordinate, $u$. We trace such modes back to ${\cal C^-}$ along past directed null geodesics through the collapsing body. Some part of the wave will be reflected back without entering it due to the greybody effect. We show that these two kind of traced back modes  yield the two temperature spectra and fluxes  subject to the aforementioned  `in' vacuum. Since the coordinate $u$  used in the `out' modes is not well defined on a horizon, estimate on how `close' we might be to ${\cal C^+}$ is given by estimating backreaction.  We argue no other reasonable choice of the `out' vacuum would give rise to any thermal spectra. Our conclusions remain valid for all non-Nariai class black holes, irrespective of the relative sizes of the two horizons.  
\end{abstract}

\newpage
\section{Introduction}\label{s1}

   It is well known that quantum mechanically a black hole can create an outward flux of particles at a temperature given by its horizon surface gravity. This astonishing phenomenon is known as the Hawking radiation and was first reported in~\cite{Hawking}, showing that a Schwarzschild or Kerr  black hole forming via a gravitational collapse would create a Planckian spectrum of particles in the asymptotic region. Soon afterwards similar results were reported with eternal black holes with a maximally analytically continued manifold and even for uniformly accelerated or Rindler observers in flat spacetimes~\cite{Unruh:1976db}.  Since then there has been a  huge endeavour to understand and explore this phenomenon in depth, we refer our reader to e.g.~\cite{DeWitt, Birrell:1982ix, Srinivasan:1998ty, Barman:2017fzh, Traschen:1999zr, Crispino:2007eb, Parker, Solodukhin:2011gn} and references therein for a vast review on various perspectives, including the quantum entanglement.

We are interested here in stationary black hole solutions of the Einstein equations  with a positive cosmological constant $\Lambda$, known as the de Sitter black holes, interesting in various ways. First, owing to the current phase of accelerated expansion of our universe, they are expected to provide nice toy models for the global structures of isolated black holes of our actual universe. Second and more important, they can model black holes formed during the inflationary phase of our universe e.g.,~\cite{Bousso:1997wi, Chao:1997em, Bousso:1999ms, Anninos:2010gh}. The most interesting qualitative feature compared to $\Lambda \leq 0$ of such black holes are the existence of the cosmological event horizon serving as a global causal boundary of the spacetime, in addition to the usual black hole horizon. The cosmological event horizon is present  in the empty de Sitter spacetime as well and it has thermal characteristics, see  e.g.~\cite{Lochan:2018pzs, Goheer:2002vf, Marolf:2010tg} and references therein for various aspects of de Sitter particle creation,  vacuum states, thermodynamics and de Sitter holography.
  
The thermodynamics and  particle creation for eternal Schwarzschild- and Kerr-de Sitter spacetimes was first studied  
in~\cite{Gibbons:1977mu} via the path integral approach. The two Killing horizons  
give rise to two temperatures, thereby making these spacetimes much more nontrivial 
and qualitatively different from their $\Lambda \leq 0$ counterparts.  Since then a lot of effort has been provided to understand various aspects of such two-temperature thermodynamics, see e.g.~\cite{Kastor:1993mj}-\cite{Bhattacharya:2017scw} and references therein.  In~\cite{Traschen:1999zr} the particle creation for eternal de Sitter black holes was studied using the Kruskal modes associated with the two horizons and the existence of non-thermal spectra was shown. The construction of the analogues of the Boulware, Hartle-Hawking and Unruh vacua for a $1+1$-dimensional Schwarzschild-de Sitter spacetime was discussed in~\cite{Choudhury:2004ph}. Study of particle creation 
including the greybody effect in higher dimensions  can be seen in~\cite{Kanti:2014dxa, Pappas:2016ovo, Kanti:2017ubd, Pappas:2017kam}.  See also~\cite{Jiang:2007mi} for a discussion on anomaly and ~\cite{Bhattacharya:2009bs} for particle creation via complex path analysis. 

However, a quantum field theoretic derivation of the particle creation for such black holes in a more realistic scenario of gravitational collapse seems to be missing. Even though the particle creation is  independent of the detail of the collapse making the construction of an eternal horizon suffice in an asymptotically flat spacetime~\cite{Birrell:1982ix}, we recall that for de Sitter black  holes we can have non-thermal spectra~\cite{Traschen:1999zr, Choudhury:2004ph} along with the thermal ones, owing to their non-trivial asymptotic structures and freedom to choose different vacuum states. Keeping such non-unique feature in mind, it is thus natural to ask in a collapse scenario for  a de Sitter black hole, what particular choices of the `in' and `out' vacua lead to the two temperature spectra and fluxes? Are these choices unique?

The present manuscript intends to fill in this gap by computing particle creation 
in a quasistationary gravitational collapse occurring in an eternal de Sitter universe, eventually to settle down to a Schwarzschild- or Kerr-de Sitter spacetime. Such modelling seems reasonable whenever the timescale of the collapse is small compared to the age of the de Sitter universe, perhaps relevant in the context of the early universe. We recall that for a black hole forming via gravitational collapse, the Kruskal coordinates making the maximal analytic extension of the manifold near its horizon has no natural meaning~\cite{Wald:1984rg}. Accordingly, we shall assign such coordinates only with the cosmological horizon.

The three causal boundaries of this spacetime will be the past and future cosmological horizon (respectively, ${\cal C^{\mp}}$) and the future black hole horizon ${\cal H^+}$, for an observer located within these horizons. We shall consider a massless, minimally coupled scalar field obeying the null geodesic approximation (e.g.~\cite{Wald:1984rg}) and will take the `in' vacuum on ${\cal C^-}$ defined with respect to the positive frequency Kruskal mode at the onset of the collapse. This choice corresponds to the fact that the Kruskal coordinates are affine generators of the null geodesics on a Killing horizon and hence represent freely falling observers. The `out' vacuum will be defined with respect to the usual positive frequency  retarded null coordinate $u$, at a radius within but `near'  ${\cal C^+}$. These out modes are then traced back to ${\cal C^-}$ along past directed null geodesics through the collapsing body. There will be greybody effect and some part of the ray will be reflected back by the potential barrier of the wave equation without entering the collapsing body. The form of these two waves are shown to be determined naturally in terms of the advanced Kruskal null coordinate  on ${\cal C^-}$,~\ref{sec.2.1}. Using these
 results, we compute the Bogoliubov coefficients and the two temperature particle spectra in~\ref{bogoliu}.  This result is further generalised to the stationary axisymmetric Kerr-de Sitter spacetime in~\ref{kds}. Finally we conclude in~\ref{disc}.   Note that the $u$-coordinate used to define the out vacuum is not well defined on a horizon. Accordingly we  provide explicit estimate at the end of~\ref{bogoliu}  on the cut-off indicating how `near' one could be to ${\cal C^+}$ to indeed consistently use it as a valid state, by estimating the backreaction. We further provide arguments in the favour of uniqueness of our choices of the vacua, as long as one is interested in getting the thermal spectra and fluxes.    \\

Even though we shall be working with individual basis modes, it will be implicitly assumed that they are linearly superposed to form suitable localised wave packets~\cite{Hawking, Traschen:1999zr, Parker}. In particular, working with basis modes instead of the wave packets is easier and is not going to affect our final results anyway.
 
Since we shall assume that the matter field obeys the null geodesic approximation, in the next section we shall discuss briefly the properties of null geodesics in the Schwarzschild-de Sitter spacetime. We shall work in $(3+1)$-dimensions with  mostly positive signature of the metric and will set $\hbar=c=k_{\rm B}=1$ throughout. 
  
\section{The metric, the null geodesics and the causal structure}\label{s2}
The Schwarzschild-de Sitter  metric  in the usual spherical polar coordinates reads
\begin{eqnarray}
ds^2=-\left(1-\frac{2M}{r}-\frac{\Lambda r^2}{3}\right)dt^2+\left(1-\frac{2M}{r}-\frac{\Lambda r^2}{3}\right)^{-1}dr^2 +r^2\left(d\theta^2+\sin^2\theta d\phi^2\right)
\label{ds1}
\end{eqnarray}
where $M$ is the mass parameter. For $3M\sqrt{\Lambda} < 1$, the above metric admits two Killing horizons,
\begin{eqnarray*}
r_{ H}=  \frac{2}{\sqrt{\Lambda}}\cos\left[\frac13 \cos^{-1}\left(3M\sqrt{\Lambda}\right)+\frac{\pi}{3} \right],~~ r_{C}=  \frac{2}{\sqrt{\Lambda}}\cos\left[\frac13 \cos^{-1}\left(3M\sqrt{\Lambda}\right)-\frac{\pi}{3} \right]
\label{ds2}
\end{eqnarray*}
where $r_H$ is the black hole event horizon (BH) and $r_C\geq r_H$ is the cosmological event horizon. For $3M\sqrt{\Lambda}=1$, these two horizons merge to $1/\sqrt{\Lambda}$, known as the Nariai limit. Beyond that limit, no black hole horizon exists and we obtain a naked curvature singularity. In other words,  a positive $\Lambda$ puts an upper bound on the maximum sizes of black holes. The surface 
gravities of these two Killing horizons are given by
\begin{eqnarray}
\kappa_H=    \frac{\Lambda (r_C+2r_H)(r_C-r_H)}{6 r_H} \qquad 
-\kappa_C=\frac{\Lambda (2r_C+r_H)(r_H-r_C)}{6 r_C}
\label{ds3}
\end{eqnarray}
where $\kappa_C>0$  and the `minus' sign in front of it ensures that the cosmological horizon  has negative surface gravity, owing to the repulsive effects of positive $\Lambda$.
Note that for $\Lambda\to 0$, we have $r_C\approx \sqrt{3/\Lambda}\to \infty$ and $r_H \approx 2M$. In that case we also recover $\kappa_H=1/2r_H=1/4M$ and $\kappa_C \approx 1/r_C\to 0$.\\

Since we shall be concerned with a massless field   propagating along null geodesic, let us first briefly discuss null geodesics in~\ref{ds1}, following the basic formalism of e.g.,~\cite{Wald:1984rg}. We rewrite \ref{ds1} as
\begin{eqnarray}
ds^2=\left(1-\frac{2M}{r}-\frac{\Lambda r^2}{3}\right)\left(-dt^2+dr_{\star}^2\right) +r^2(r_{\star})d\Omega^2
\label{ds4}
\end{eqnarray}
where $r$ as a function of the tortoise coordinate $r_{\star}$ is understood,  given by
\begin{eqnarray}
r_{\star}=\int \left(1-\frac{2M}{r}-\frac{\Lambda r^2}{3}\right)^{-1}\,dr=\frac{1}{2\kappa_H}\ln \left(\frac{r}{r_H}-1\right) -\frac{1}{2\kappa_C} \ln \left(1-\frac{r}{r_C}\right) +\frac{1}{2\kappa_u}\ln \left(\frac{r}{r_H+r_C}+1\right) 
\label{ds5}
\end{eqnarray}
where
%
$\kappa_u= (M/r_u^2-\Lambda r_u/3)$
%
corresponds to the `surface gravity' of the unphysical negative root, $r_u=-(r_H+r_C)$, of $g_{tt}=0$. It is easy to see by expanding the log's in~\ref{ds5} that in the limit $\Lambda \to 0$, i.e. $r_C\to \infty$,
we recover the Schwarzschild limit, $r_{\star}\approx r+ 2M \ln (r/2M-1)$.

Since both $\kappa_H$ and $\kappa_C$ are positive, \ref{ds5} shows  that $r_{\star}\to \mp \infty$ as $r\to r_H, r_{C}$, respectively. 
We define the retarded and the advanced null coordinates $u=t-r_{\star}$ and $v=t+r_{\star}$ to rewrite~\ref{ds4} as 
\begin{eqnarray}
ds^2=-\left(1-\frac{2M}{r}-\frac{\Lambda r^2}{3}\right)\,dudv +r^2(u,v)d\Omega^2
\label{ds7}
\end{eqnarray}
and consider incoming and outgoing radial null geodesics, $u^a$. Due to the time translation symmetry existing in between the two horizons, we have the conserved energy, $E=(1-2M/r-\Lambda r^2/3)\frac{dt}{d\lambda}$, where $\lambda$ is an affine parameter along the geodesic. Using this along with the null geodesic dispersion relation, $u_au^a=0$, we get for radially outgoing and incoming future directed geodesics,
\begin{eqnarray}
\frac{dr}{d\lambda}=+E~~({\rm outgoing})\qquad \frac{dr}{d\lambda}=-E~~({\rm incoming}).
\label{ds8}
\end{eqnarray}
The above equations yield,
\begin{eqnarray}
\frac{du}{d\lambda}=0~~({\rm outgoing})\qquad \frac{dv}{d\lambda}=0~~({\rm incoming}).
\label{ds9}
\end{eqnarray}
Using these it is further easy to find out the variations of $u$ and $v$ respectively along incoming and outgoing geodesics, 
\begin{eqnarray}
u_{\rm in}=-2r_{\star}+u_0~\qquad v_{\rm out}=2r_{\star}+v_0,
\label{ds10}
\end{eqnarray}
where $u_0$ and $v_0$ are integration constants.
Recalling $r_{\star}\to \mp \infty$ as $r\to r_H, r_{C}$, we obtain
\begin{eqnarray}
u_{\rm in}(r_H)\to \infty, \quad u_{\rm in}(r_C) \to -\infty,~~{\rm and }~~v_{\rm out}(r_H)\to -\infty, \quad v_{\rm out}(r_C)\to \infty
\label{ds11}
\end{eqnarray}
In other words, we may identify the past/future segments of the Killing horizons from the negative/positive infinities of the null coordinates. For example, 
the first of the above equations represent the future black hole horizon $({\cal H^+})$ whereas the second represent the past cosmological horizon $({\cal C^-})$
and so on. 

\ref{ds8} gives $r_{\rm out}= E\lambda +r_0$ and $r_{\rm in}= -E\lambda +r_0'$. We set both the integration constants $r_0$ and $r_0'$ to $r_H$
for convenience. From~\ref{ds5}, \ref{ds10} we now have near the horizons
\begin{eqnarray}
(u-u_0)_{\rm in}\vert_{r\to r_H}\approx -\frac{1}{\kappa_H}\ln \left(-\frac{E\lambda}{r_H}\right), \quad (u-u_0)_{\rm in}\vert_{r\to r_C} \approx \frac{1}{\kappa_C} \ln \left(1-\frac{(-E\lambda+r_H)}{r_C}\right)\nonumber\\
(v-v_0)_{\rm out}\vert_{r\to r_H}\approx \frac{1}{\kappa_H}\ln \frac{E\lambda}{r_H}, \quad (v-v_0)_{\rm out}\vert_{r\to r_C}\approx -\frac{1}{\kappa_C} \ln \left(1-\frac{(E\lambda+r_H)}{r_C}\right)
\label{ds12}
\end{eqnarray}
which, via~\ref{ds11}, fix the values of $\lambda$ on the horizons. In particular, we always have $\lambda=0$ on both past and future segments of BH, $r=r_H$.

Using~\ref{ds5}, we can cast the metric~\ref{ds7} into two alternative forms,
\begin{eqnarray}
ds^2=-\frac{2M}{r}\left(1-\frac{r}{r_C}\right)^{1+\frac{\kappa_H}{\kappa_C}}\left(1+\frac{r}{r_H+r_C}\right)^{1-\frac{\kappa_H}{\kappa_u}}e^{\kappa_H (v-u) }\,dudv+r^2d\Omega^2
\label{ds13}
\end{eqnarray}
and
\begin{eqnarray}
ds^2=-\frac{2M}{r}\left(\frac{r}{r_H}-1\right)^{1+\frac{\kappa_C}{\kappa_H}} \left(1+\frac{r}{r_H+r_C}\right)^{1+\frac{\kappa_C}{\kappa_u}} e^{\kappa_C (u-v)}\,dudv+r^2d\Omega^2,
\label{ds14}
\end{eqnarray}
Note the first and the second have  no coordinate singularities at $r=r_H,\,r_C$, respectively. We now define the Kruskal null coordinates for the black hole and the cosmological horizon
as, e.g.~\cite{Traschen:1999zr},
\begin{eqnarray}
U_H=-\frac{1}{\kappa_H}e^{-\kappa_H u},\qquad V_H=\frac{1}{\kappa_H}e^{\kappa_H v}; \qquad
U_C=\frac{1}{\kappa_C}e^{\kappa_C u},\qquad V_C=-\frac{1}{\kappa_C}e^{-\kappa_C v}.
\label{ds15}
\end{eqnarray}

We note from \ref{ds11} that $V_C \to 0$ on the future cosmological horizon. Thus on the past cosmological horizon, we must have $-\infty <V_C\leq 0 $. This will be useful for our future purpose.\\

In terms of the Kruskal coordinates defined in \ref{ds15},  the  metrics~\ref{ds13}, \ref{ds14} respectively become
\begin{eqnarray}
ds^2=-\frac{2M}{r}\left(1-\frac{r}{r_C}\right)^{1+\frac{\kappa_H}{\kappa_C}} \left(1+\frac{r}{r_H+r_C}\right)^{1+\frac{\kappa_C}{\kappa_u}}\, dU_H dV_H+r^2d\Omega^2
\label{ds16}
\end{eqnarray}
and  
\begin{eqnarray}
ds^2=-\frac{2M}{r}\left(\frac{r}{r_H}-1\right)^{1+\frac{\kappa_C}{\kappa_H}} \left(1+\frac{r}{r_H+r_C}\right)^{1+\frac{\kappa_C}{\kappa_u}}\, dU_C dV_C+r^2d\Omega^2.
\label{ds17}
\end{eqnarray}
As a consistency check  for \ref{ds17} we let $\Lambda \to 0$, implying $r_C\to \infty$, $\kappa_C\to 0$, $\kappa_C=\kappa_u$ and $r_H\to 2M$. Also in this limit, we have from \ref{ds15}, $U_C\approx \kappa_C^{-1}+u$ and $V_C\approx -\kappa_C^{-1} +v $. Using these it is easy to see that  the above metric  reduces to
the Schwarzschild spacetime.  

Let us now consider an outgoing null geodesic at $r \to r_C$. Using~\ref{ds14}, we rewrite the expression for the conserved energy, $E=(1-2M/r-\Lambda r^2/3) \frac{dt}{d\lambda}$ as 
%
$$E=\frac{2M}{r_C}\left(\frac{r_C}{r_H}-1\right)^{1+\frac{\kappa_C}{\kappa_H}}\left(1+\frac{r_C}{r_H+r_C}\right)^{1+\frac{\kappa_C}{\kappa_u}}e^{-\kappa_C (v-u) }\, \frac{dt}{d\lambda},$$
%
which after using $t=(u+v)/2$, recalling that $u=~{\rm const.}$ along outgoing geodesics and~\ref{ds15}, gives
\begin{eqnarray}
\lambda=\lambda_0+ \left[\frac{M}{r_C E}\left(\frac{r_C}{r_H}-1\right)^{1+\frac{\kappa_C}{\kappa_H}}\left(1+\frac{r_C}{r_H+r_C}\right)^{1+\frac{\kappa_C}{\kappa_u}}e^{\kappa_C u }\right] V_{C}
\label{ds19}
\end{eqnarray}
where $\lambda_0$ is an integration constant.
The above equation shows that $V_C$ (or any coordinate proportional to it with a constant coefficient of proportionality) would be an affine parameter along the outgoing null geodesics on $r=r_C$. Likewise, it turns out that $U_C$ is an affine 
parameter along the incoming null geodesics there.

Similar conclusions hold on the black hole event horizon, with the respective Kruskal coordinates $V_H$ and $U_H$. 

\begin{figure}[h]
\includegraphics[width=6cm]{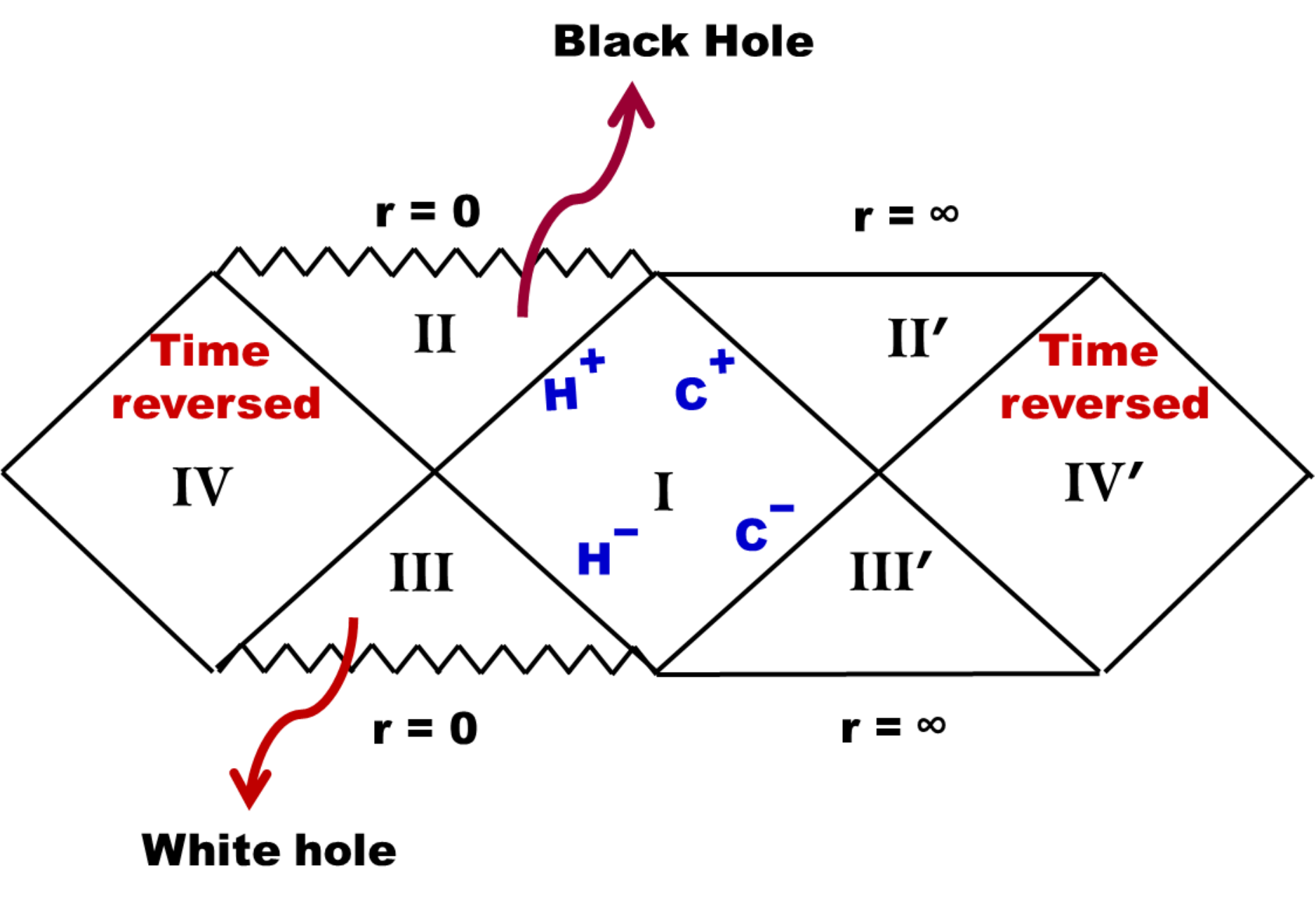}
\centering
\caption{\it The Penrose-Carter diagram for the Schwarzschild-de Sitter spacetime~\ref{ds1}, e.g.~\cite{Traschen:1999zr, Choudhury:2004ph}. Each point should be understood as tangent to a 2-sphere centred at $r=0$. The spacetime could be indefinitely analytically continued further to the left and right by adding further mass points. A physical observer is usually taken to be located in the diamond shaped region, Region I,  held within  ${\cal H }^{\pm}$ and ${\cal C}^{\pm}$.}
\centering
\label{fig1'}
\end{figure}

\ref{fig1'} depicts the Penrose-Carter diagramm of \ref{ds1}, e.g.~\cite{Traschen:1999zr, Choudhury:2004ph}. On the past black hole horizon ${\cal H}^-$, a particle/field  could be outgoing only, as opposed to the future black hole horizon,  ${\cal H}^+$, where it can be ingoing only. Likewise on ${\cal C}^-$, we have incoming trajectories  only, as opposed to ${\cal C}^+$.   \\
\begin{figure}[h]
\includegraphics[width=3cm]{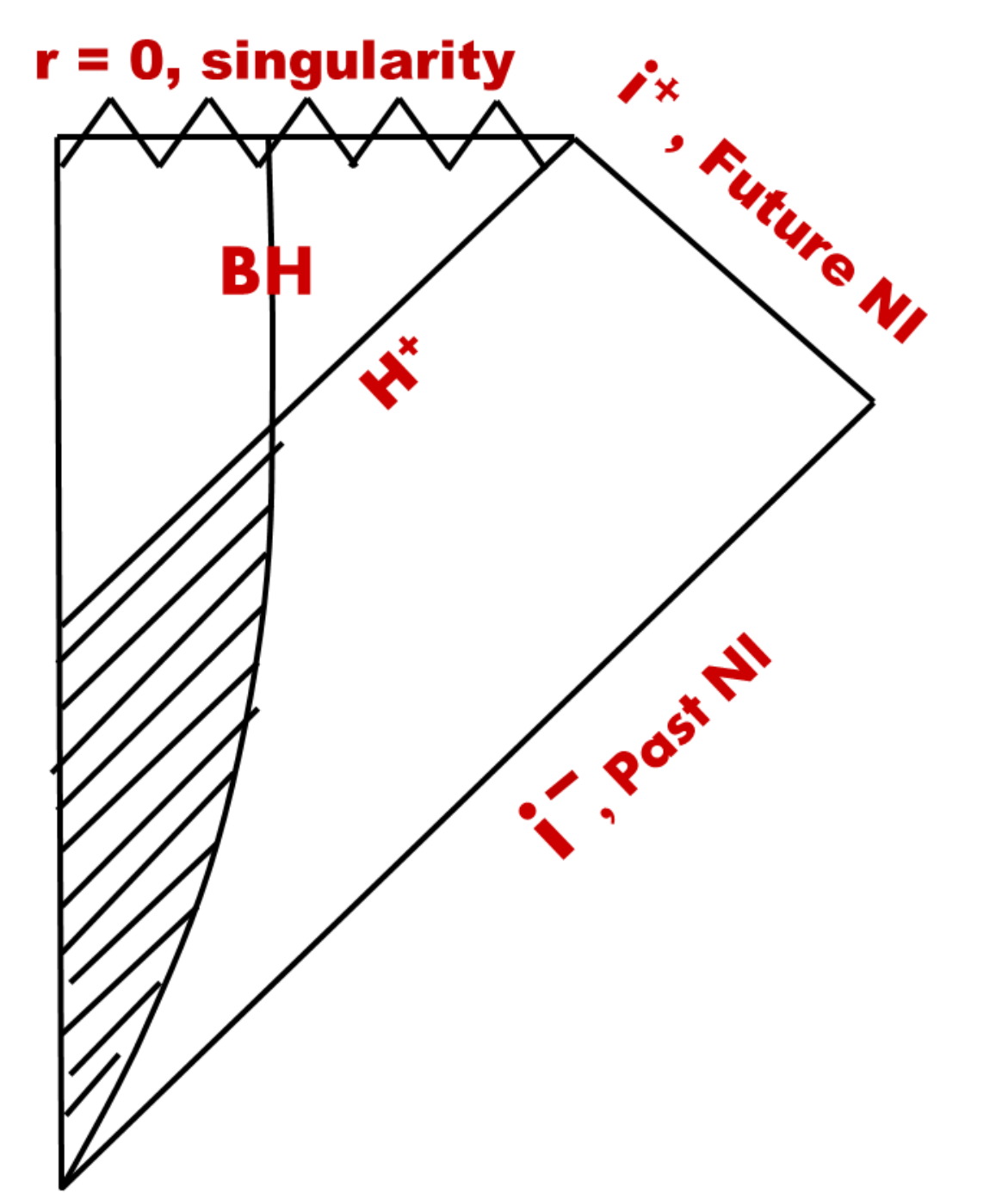}
\includegraphics[width=3cm]{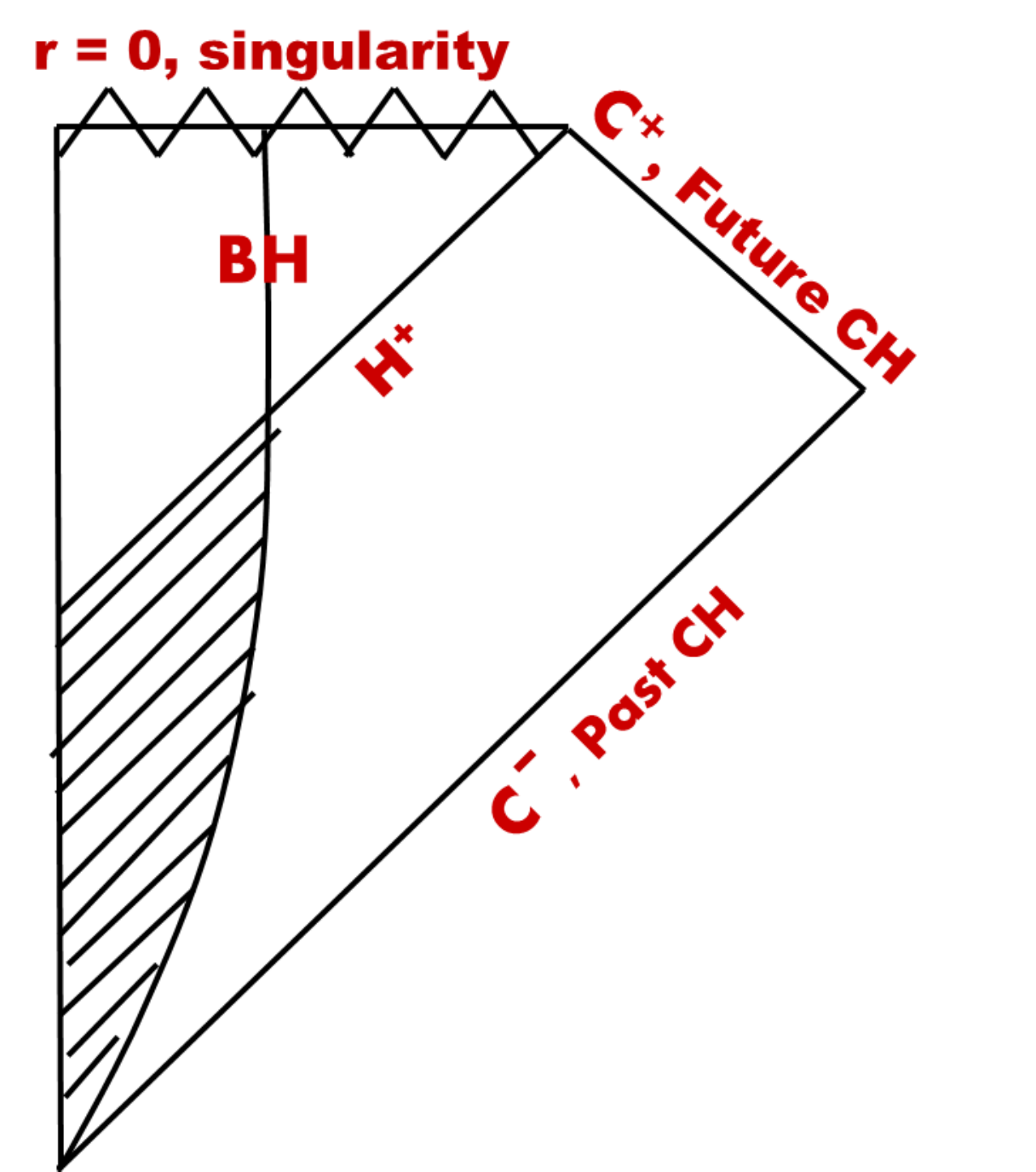}
\centering
\caption{\it The Penrose-Carter diagrams for gravitational collapses forming  black holes respectively a) in an asymptotic flat spacetime, e.g.~\cite{Wald:1984rg}, and b) within the de Sitter event horizons. The shady region denotes the collapsing body, entering the future black hole horizon, ${\cal H}^+$.}
\centering
\label{fig2}
\end{figure}

In this work, we are particularly interested in non-eternal black holes forming via a gravitational collapse in a de Sitter spacetime. This means that we take the de Sitter space to be eternal and 
consider a gravitational collapse occurring within it, to form a black hole. In that case  the white hole horizon ${\cal H}^-$ will be absent.  In other words in the collapse scenario the Kruskal coordinate associated with the black hole, \ref{ds16}, has no meaning and {\it only} such coordinates with the cosmological horizon, \ref{ds17}, is relevant.

The first of \ref{fig2}  depicts a gravitational collapse forming a black hole in an asymptotically flat spacetime. Massless modes originating from the past null infinity ${\cal I}^-$ enter the collapsing body, gets scattered and propagates to the future null infinity, ${\cal I}^+$ before the horizon forms. The second of \ref{fig2} represents similar process occurring inside the past and future cosmological event horizons.  Precisely, modes incoming from ${\cal C}^-$ eventually enters the collapsing body, and gets scattered off to ${\cal C}^+$. The `in' vacuum 
would correspond to the appropriate positive energy mode functions incoming at ${\cal C}^-$, whereas the `out' vacuum will correspond to the scattered positive frequency outgoing modes near ${\cal C}^+$.  We shall be more precise about them in what follows. 
\section{The field equation and particle creation} \label{sec.2}
\subsection{The mode functions}\label{sec.2.1}
We assume that the collapse process is quasistationary i.e., outside the collapsing body we can approximate the spacetime 
by~\ref{ds1}.
Let us then consider a massless, minimally coupled test scalar field $\psi$, propagating in this background satisfying the Klein-Gordon equation,
$$\nabla^a \nabla_a \psi=0$$
For any two solutions $f_1$ and $f_2$, the Klein-Gordon inner product reads,
\begin{eqnarray}
(f_1,f_2)=i\int d\Sigma \, \left(f_1^{\star} \nabla_a f_2 -f_2 \nabla_a f_1^{\star} \right) n^a 
\label{ds20''}
\end{eqnarray}
where the integration is done over any suitable  hypersurface and $n^a$ is the unit normal to it. 

Employing the usual variable separation for individual mode functions, $\psi_{\omega l m}(t,r, \theta, \phi)=\frac{R_{\omega l}(r,t) }{r}  Y_{l m}(\theta, \phi)$, we get the familiar wave equation with an effective potential, 
\begin{eqnarray}
\left(-\frac{\partial^2 }{\partial t^2} +\frac{\partial^2}{\partial r_{\star}^2 }\right)R_{\omega l}-\left(1-\frac{2M}{r}-\frac{\Lambda r^2}{3}\right)\left(\frac{l(l+1)}{r^2}+\frac{2M}{r^3}-\frac{2\Lambda}{3}\right)R_{\omega l}=0
\label{ds20}
\end{eqnarray}
where $r_{\star}$ is given by~\ref{ds5}. If we further insert the customary time dependence $e^{\pm i \omega t}$, the above equation takes the form of the Schr\"{o}dinger equation with an effective potential vanishing on the two  horizons with a maximum in between.  Thus analogous to the ordinary scattering problem, a wave  incoming from ${\cal C^-}$ will be split into two parts upon incident on this effective potential barrier -- the transmitted wave will propagate inward whereas the reflected wave will turn back towards the cosmological horizon.
 
In terms of the retarded and the advanced null coordinates $u$ and $v$, the mode functions  $R_{\omega l}$ on or in an infinitesimal neighbourhood of any of the horizons 
become plane waves
\begin{eqnarray}
 R_{\omega l} \sim  e^{-i\omega u}, \, e^{-i\omega v}, 
\label{ds20'}
\end{eqnarray}
along with their negative frequency counterparts. 

As we mentioned earlier, only the Kruskal coordinates for the cosmological event horizon, \ref{ds17}, should be relevant in the collapse scenario. In terms of the
Kruskal timelike $(T_C)$ and radial $(R_C)$ coordinates,  $U_C=T_C-R_C$ and  $V_C=T_C+R_C$, \ref{ds17} reads 
\begin{eqnarray}
ds^2=\frac{2M}{r}\left(\frac{r}{r_H}-1\right)^{1+\frac{\kappa_C}{\kappa_H}} \left(1+\frac{r}{r_H+r_C}\right)^{1+\frac{\kappa_C}{\kappa_u}}\, (-dT_C^2+dR_C^2) +r^2d\Omega^2.
\label{ds21}
\end{eqnarray}
Using the above and the last two of~\ref{ds15}, the field equation \ref{ds20}
takes the form as $r\to r_C$,
\begin{eqnarray}
\left[-\frac{\partial^2 }{\partial T_c^2} + \frac{\partial^2 }{\partial R_c^2}\right]R^K_{\omega l}-\frac{2M}{r}\left( \frac{r}{r_H}-1\right)^{\frac{\kappa_C}{\kappa_H}+1}\left(1+\frac{r}{r_H+r_C}\right)^{\frac{\kappa_C}{\kappa_u}+1}\left( \frac{l(l+1)}{r^2}+\frac{2M}{r^3}-\frac{2\Lambda}{3}\right)R^K_{\omega l}=0
\label{ds22}
\end{eqnarray}
The  solutions with positive frequency (with respect to the Kruskal timelike coordinate $T_c$) basis modes read
\begin{eqnarray}
R^K_{\omega l} \sim e^{-i(\omega T_c-k R_c)},\,  e^{-i(\omega T_c+kR_c)},
\label{ds23}
\end{eqnarray}
where 
$$\omega^2 = k^2 +\frac{2M}{r_C}\left( \frac{r_C}{r_H}-1\right)^{\frac{\kappa_C}{\kappa_H}+1}\left(1+\frac{r_C}{r_H+r_C}\right)^{\frac{\kappa_C}{\kappa_u}+1}\left( \frac{l(l+1)}{r_C^2}+\frac{2M}{r_C^3}-\frac{2\Lambda}{3}\right) $$

However, in order that $\psi $ propagates along null geodesic, we must have $k^2$ to be much greater than the second term appearing on the right  hand side of the above equation,  so that the phase indeed satisfies the dispersion relation, $\omega^2-k^2 \approx 0$. 
We thus make the following mode expansion for the field operator $\psi$ at the {\it onset} of the collapse, compatible with the null geodesic approximation, localised around   ${\cal C}^-$ and incoming there
\begin{eqnarray}
\psi (x) = \int_{0}^{\infty} \frac{d\omega}{\sqrt{2\pi}\, \sqrt{2 \omega}} \sum_{lm}  \left[a_{\omega l m} \,e^{-i \omega V_C}\,\frac{Y_{lm} (\theta, \phi)}{r_C} +a_{\omega l m}^{\dagger} e^{i \omega V_C}\,\frac{Y^{\star}_{lm} (\theta, \phi)}{r_C}   \right]
\label{ds24}
\end{eqnarray}
  The orthonormality of the basis modes in~\ref{ds24} can easily be verified using \ref{ds20''} on a $T_c={\rm const.}$ hypersurface in a neighbourhood of  ${\cal C^-}$ ($n^a$ in \ref{ds20''} is taken to be a timelike unit vector along $\partial_{T_c}$). 
The creation and annihilation operators satisfy the canonical commutation relations,
\begin{eqnarray}
[a_{\omega l m}, a^{\dagger}_{\omega' l', m'}]= \delta (\omega-\omega') \delta_{ll'}\delta_{m m'},~~[a_{\omega l m}, a_{\omega' l' m'}]=0=[a^{\dagger}_{\omega l m}, a^{\dagger}_{\omega' l' m'}]
\label{ds25}
\end{eqnarray}
and the `in' vacuum is given by, 
$$a_{\omega l m}|0,{\rm in}\rangle=0.$$
Note that unlike the asymptotic flat spacetime~\cite{Hawking}, we cannot take the ordinary $v$-modes to define the `in' vacuum here. This is because  $V_C$ and not $v$ is a null geodesic generator of the cosmological horizon (c.f., discussions below \ref{ds17}). In other words, choosing the coordinate of freely falling observers  here seems to naturally carry the intuitive notion that the vacuum should be the state most compatible  with the spacetime geometry. 
\\

Let us now consider modes propagating inward after emanating from ${\cal C}^-$ and entering the collapsing body in the second of \ref{fig2}. In order to describe this dynamics, we shall 
take the usual $(u,\,v)$ modes. Precisely, we shall consider incoming $v$-modes entering the body, getting scattered by the centrifugal barrier at $r=0$, off to ${\cal C}^+$. However, since the $(u,v)$-coordinates are not well defined on the horizons, \ref{ds7},  we cannot have a physical vacuum on ${\cal C^+}$ for modes written in these coordinates, similar to the asymptotic flat spacetimes~\cite{Birrell:1982ix}. Accordingly, 
we shall imagine intercepting such modes by a static observer on their way to ${\cal C^+}$ at a point {\it close to, but not on } ${\cal C^+}$.
Following~\cite{Hawking} (also e.g.~\cite{Traschen:1999zr, Parker})  we shall then trace 
these modes  back onto ${\cal C}^-$ through the collapsing body via the null geodesic approximation and then will compute the Bogoliubov coefficients with respect to~\ref{ds24}.  We can thus specify the  `out' mode functions near ${\cal C^+}$ and on ${\cal H^+}$ as    
\begin{eqnarray}
\psi(x)=  \int_{0}^{\infty} \frac{d\omega}{\sqrt{2\pi}} \sum_{lm}\left[b_{\omega l m }\, \zeta_{\omega l} (r) {\frac{e^{-i \omega u}}{\sqrt{2\omega}}}\,\frac{Y_{lm} (\theta, \phi)}{r} +b_{\omega l m}^{\dagger}\, \zeta_{\omega l}^{\star} (r) {\frac{e^{i \omega u}}{\sqrt{2 \omega}}}\, \frac{Y^{\star}_{lm}(\theta, \phi)}{r} +  c_{\omega l m } q_{\omega l m}+ c^{\dagger}_{\omega lm }q^{\dagger}_{\omega l m}\right] 
\label{ds25'}
\end{eqnarray}
where $q_{\omega l m}$'s are some complete ingoing mode functions localised on ${\cal H^+}$ and the operators $b_{\omega l m }$ and $c_{\omega l m }$ separately satisfy commutation relations analogous to~\ref{ds25}. The functions $\zeta$ and $\zeta^{\star}$ smoothly reach unity as $r\to r_C$, by \ref{ds20}. A power series solution of them can easily be found in the neighbourhood of ${\cal C^+}$, perhaps useful for determining the exact renormalised energy-momentum tensor but we shall not go into that here.  Also, the inner product between the mode functions in \ref{ds25'} vanish, as they are mutually disconnected. Let us now imagine a ray, which after getting scattered within the collapsing object, leaves it just before the black hole horizon forms.  Then, \\

\noindent
(i) For an outgoing mode $\sim e^{-i\omega u}$,  by~\ref{ds11}, \ref{ds12}, the advanced null coordinate $v$ would diverge logarithmically on the future black hole horizon. This mode would eventually reach ${\cal C^+}$ with high value of the retarded coordinate $u$. We trace this back to ${\cal C^-}$ along a 
past directed  null geodesic through the collapsing body. Some part of the wave during this process will be scattered back towards ${\cal C^+}$ by the effective potential barrier of~\ref{ds20} before entering the body.\\

\noindent
(ii) As we trace the ray back through the collapsing body, the outgoing ray becomes incoming $\sim e^{-i\omega v}$ and accordingly by continuity, the phase of the mode function is given by $v\equiv v(u)=-\frac{1}{\kappa_H}\ln (-C \lambda)+v_0$,~\ref{ds12}, where $C>0$ and $v_0$ are constants.   \\

\noindent
(iii) Since $\lambda \to 0^-$, ~\ref{ds11}, \ref{ds12}, such modes would always propagate along null geodesics. However, when such a mode is traced back to ${\cal C^-}$ along a past directed outgoing null geodesic, the functional form of the  affine parameter $\lambda$ will be changing in a continuous manner.  What is its final form on ${\cal C^-}$? Certainly, since such modes are past directed and outgoing on ${\cal C^-}$, it would be given there by the affine parameter along the outgoing null geodesic generator of the cosmological event horizon, i.e. $\lambda=C'( V_C-V_0)$, where $C'>0$ and $V_0$ are constants (cf., discussions after~\ref{ds19}) with $(V_C-V_0)$ being negative and close to zero.  Putting these all in together and using~\ref{ds20'}, we find  
the following expressions for the positive frequency mode functions, say $p_{\omega l m}$, on ${\cal C^-}$ {\it after} the ray tracing, in terms of the advanced null coordinate $V_C$ : 
\begin{eqnarray}
 p^{(1)}_{\omega l m} &=& {\frac{e^{-i \omega v }}{\sqrt{2\pi} \sqrt{2 \omega}}}\,\frac{Y_{lm} (\theta, \phi)}{r_C}={\frac{e^{ \frac{i\omega}{\kappa_C} \ln (-\kappa_C V_C) }}{\sqrt{2\pi}\sqrt{2 \omega}}}\,\frac{Y_{lm} (\theta, \phi)}{r_C}~~({\rm modes~that~did~not~enter~the~collapsing~body} )\nonumber \\
 p^{(2)}_{\omega l m} &=& {\frac{e^{ \frac{i\omega}{\kappa_H}\ln (C_0 (V_C-V_0))  }}{\sqrt{2\pi} \sqrt{2 \omega}}}\,\frac{Y_{lm} (\theta, \phi)}{r_C}~~({\rm entered~and~got~scattered~just~prior~to}~{\cal H^+}~{\rm formed})
\label{ds26}
\end{eqnarray}
where in the first line we have used \ref{ds15} and $C_0<0$ is a constant whose explicit form is not necessary for our present purpose. Note that $(V_C-V_0)$ has to be less than or equal to zero in $p^{(2)}_{\omega l m}$. In other words $V_C=V_0$ should denote the last ray that could escape to ${\cal C^+}$ just before ${\cal H^+}$ forms. Replacing $V_C$ by $v$, the null geodesic generator of ${\cal I^-}$, in the second of the above equations recovers the result of the asymptotically flat spacetime~\cite{Hawking}. There could be some additional global phases in the above modes  owing to the scattering by the effective potential  barrier of \ref{ds20}. However, such phases will not affect our computations anyway. 

We must also note that the two traced back waves  in \ref{ds26} are located in {\it disjoint regions} on ${\cal C^-}$ as follows. Let us  trace back a single positive frequency, outgoing  mode function with a given value of the retarded coordinate $u$  in \ref{ds25'}. This splits into the aforementioned two segments, both essentially propagating with the speed of light. Thus the wave that enters the collapsing body, having traveled a longer `distance' compared to the other one, would take longer to reach ${\cal C^-}$. Thus when finally they are traced back to ${\cal C^-}$, they would correspond to two different values of the advanced null coordinate $V_C$. Conversely, if a wave starts from ${\cal C^-}$ with some given value of $v$ or $V_C$, by the time the part of the wave that enters the collapsing body reaches the static observer located near ${\cal C^+}$, the part  that did not enter would already have crossed him/her. Thus any two {\it traced back}  rays $p^{(1)}_{\omega l m}$ and $p^{(2)}_{\omega l m}$  to ${\cal C^-}$  with the same $v$ in \ref{ds26} must be disjoint there. 

The mode $p^{(1)}_{\omega l m}$ corresponds to the classical greybody effect and in the asymptotically flat black hole spacetimes e.g.~\cite{Hawking, Traschen:1999zr, Parker}, their vacuum   coincides  with the `in' vacuum. For our case however, \ref{ds26} shows that they would indeed suffer non-trivial redshift with respect to the Kruskal `in' modes hence would also give particle creation effects. Finally, as we have argued above, since the traced back modes $p^{(1)}_{\omega l m}$ 
and $p^{(2)}_{\omega l m}$ propagate to regions of disjoint supports on ${\cal C^-}$, we may compute the Bogoliubov coefficients associated with them independently with respect to \ref{ds24}.  

Being equipped with all these, we shall now compute the two-temperature spectra of created particles for the Schwarzschild-de Sitter spacetime below.

\subsection{The Bogoliubov coefficients and the particle spectra}\label{bogoliu}
The techniques for the computation of the Bogoliubov coefficients and the spectra of the created particles essentially parallel to that of the asymptotically flat spacetime, e.g.~\cite{Hawking,  DeWitt, Traschen:1999zr, Parker}. We shall mention only some key steps below for the sake of completeness.

We first note from \ref{ds26} that the two modes are seemingly energetically different and hence the vacuum  associated with them would also be different.  Next, using \ref{ds26}  we compute the Bogoliubov  relations between \ref{ds24} and \ref{ds25'} on ${\cal C^-}$. Since $q_{\omega l m}$'s have support only on ${\cal H^+}$, they are not going to affect our computations. 
Using the integral representation~\cite{Copson},
$$\int_{0}^{\infty}dp\, p^{\alpha-1}\,e^{-ip}=e^{-\frac{i \pi \alpha}{ 2}}\, \Gamma (\alpha)~~({\rm Re}(\alpha)>0)$$
and \ref{ds20''}, we find for the  annihilation operator associated with  second mode function in~\ref{ds26}
\begin{eqnarray}
b^{(2)}_{\omega l m}=-\sqrt{f_{\omega}}\,\frac{i(-C_0)^{\frac{i\omega}{\kappa_H}}}{2\pi \sqrt{\omega}}\Gamma\left(\frac{i\omega}{\kappa_H}+1\right) \int_{0}^{\infty} d\omega' (\omega')^{-\frac{i\omega}{\kappa_H}-\frac12} \left[e^{\frac{\pi \omega}{2 \kappa_H}+i\omega' V_0} a_{\omega' l m}+ e^{-\frac{\pi \omega}{2\kappa_H}-i\omega' V_0} a^{\dagger}_{\omega' l m}\right]
\label{ds27}
\end{eqnarray}
Using now
\begin{eqnarray}
\int_{0}^{\infty}dp\, p^{-1\pm ix} = 2\pi \, \delta (x),\qquad |\Gamma(1+ix)|^2 = \frac{\pi x}{\sinh\pi x}
\label{formulae}
\end{eqnarray}
and recalling $C_0$ is negative and real it is easy to show that
\begin{eqnarray}
[b^{(2)}_{\omega_1 l m}, b^{(2)^{\dagger}}_{\omega_2 l' m'}]= f_{\omega_1 }\,\delta(\omega_1-\omega_2)\delta_{ll'}\delta_{mm'}, \quad [b^{(2)}_{\omega_1 l m}, b^{(2)}_{\omega_2 l' m'}]=0=[b^{(2)^{\dagger}}_{\omega_1 l m}, b^{(2)^{\dagger}}_{\omega_2 l' m'}]
\label{ds28}
\end{eqnarray}
where we have also used the fact that $\omega_1, \omega_2, \kappa_H \geq 0$. Also $f_{\omega }<1$ is the usual greybody function introduced in order to avoid overcompleteness. For the mode $p^{(1)}_{\omega l m}$ we would have
\begin{eqnarray}
b^{(1)}_{\omega l m}=-\sqrt{1-f_{\omega }}\,\frac{i(\kappa_C)^{\frac{i\omega}{\kappa_C}}}{2\pi \sqrt{\omega}}\Gamma\left(\frac{i\omega}{\kappa_C}+1\right) \int_{0}^{\infty} d\omega' (\omega')^{-\frac{i\omega}{\kappa_C}-\frac12} \left[e^{\frac{\pi \omega}{2\kappa_C}} a_{\omega' l m}+ e^{-\frac{\pi \omega}{2\kappa_C}} a^{\dagger}_{\omega' l m}\right]
\label{ds29}
\end{eqnarray}
It is easy to see that the above satisfies commutation relations analogous to \ref{ds28}. \\

We note that the existence of the two annihilation operators $b^{(1)}$ and $b^{(2)}$ implies the existence of two `out' vacua
for the two kinds of modes we discussed. Evidently this is not in any sense  doubling of degrees of freedom, for they just correspond to the two scattered part of the wave which are energetically different for having underwent different redshifts. 

The particles created by the black hole in a given eigenmode is  : $N_{\omega l m}=\langle 0,{\rm in}| b^{(2)\dagger}_{\omega l m} \, b^{(2)}_{\omega l m} |0,{\rm in}\rangle$ (no sum on $\omega, l, m$) which by \ref{ds27}, turns out to be divergent showing the total number of created particles is infinite. However, the particles created per unit Kruskal time is finite as follows. We instead evaluate $\lim_{\epsilon \to 0}\langle 0,{\rm in}| b^{(2)\dagger}_{\omega+\epsilon\, l m} \, b^{(2)}_{\omega l m} |0,{\rm in}\rangle$ and use
$$\delta(0)= \frac{1}{2\pi}   \lim_{T_C \to \infty} \,\left(\lim_{\epsilon \to 0}\int_{-T_C/2}^{+T_C/2} dT_C \,e^{\pm i \epsilon T_C}\right)= \lim_{T_C \to \infty}\frac{T_C}{2\pi}, $$
to find the  thermal spectrum of created particles by the black hole  per unit cosmological Kruskal time in the vicinity of  ${\cal C^+}$, 
\begin{eqnarray}
n^{H}_{\omega }= \lim_{T_C\to \infty} \frac{N_{\omega }}{T_C} = \frac{1}{2\pi}\frac{f_{\omega }}{e^{\frac{2\pi \omega}{\kappa_H}}-1}
\label{ds30}
\end{eqnarray}
Likewise for the particles created by the cosmological horizon we obtain 
\begin{eqnarray}
n^{C}_{\omega }= \frac{1}{2\pi}\frac{1-f_{\omega }}{e^{\frac{2\pi \omega}{\kappa_C}}-1}
\label{ds31}
\end{eqnarray}

A couple of points should follow here.
First, we could have chosen the cosmological Kruskal modes itself to describe the collapse. In that case, it is easy to see that the first traced back ray in~\ref{ds26} simply behaves as $\sim e^{-i \omega V_C}$ whereas the second as $e^{\frac{i \omega}{\kappa_C}\,(-C_0(V_C-V_0))^{\kappa_C/\kappa_H}}$. The first would give no particle creation for the cosmological horizon whereas the second would predict a non-thermal spectrum for the black hole. Even though there is no computational inconsistency in it,  we note that the Kruskal coordinates carry natural physical interpretation with its respective horizon only. Thus using a mode written in terms of the cosmological Kruskal coordinate on the black hole event horizon (while doing the tracing back the modes)  may not carry any natural physical meaning. Likewise we could have defined the `out' vacuum with respect to  the Kruskal mode $\sim e^{-i\omega U_C}$ on ${\cal C^+}$. Since both `in' and `out' vacuum on the cosmological horizon will correspond to freely falling observer's coordinate, there will be no particle creation for the cosmological horizon in this vacuum. Also as we argued above, such modes are not much meaningful to be used to describe the collapse dynamics and ray tracing.   Putting these all in together, it seems that our choice of the `out' vacuum is the only reasonable choice we could make, given that the `in' vacuum is unique in the present scenario.

Second,  we note from \ref{ds25'} that the derivatives of the phases will dominate the other spatial derivatives by virtue of the null geodesic approximation. Using then \ref{ds27} and the regularised sum $\sum_{0}^{\infty}(2l+1)=1/12$, we obtain after normal ordering the leading expression for the flux of particles created by the black hole at the position of our stationary observer, say $r=r_0 < r_C$,
\begin{eqnarray}
\int r_0^2 d\Omega^2\langle 0, {\rm in}| T_{t}{}^r| 0, {\rm in}\rangle \approx \frac{|\zeta (r_0)|^2}{24 \pi }\int_{0}^{\infty}
\frac{d\omega\, \omega\, f_{\omega} } {e^{\frac{2\pi \omega}{\kappa_H }} -1} 
\label{flux4}
\end{eqnarray}
showing the existence of a thermal flux. Likewise by using \ref{ds29} we obtain the same for the cosmological horizon, replacing $\kappa_H$ by $\kappa_C$ and $f_{\omega}$ by $(1-f_{\omega})$. 

Third, recalling that a $u$-mode is unphysical on a Killing horizon for it yields divergent expectation value of the energy-momentum tensor e.g.~\cite{Birrell:1982ix},  we would like to make a heuristic estimate on how  `close' one could be to ${\cal C^+}$ to consistently work with our choice of the `out' vacuum.  We have at $r=r_0$ for the particles created by the black hole,
$$\langle 0, {\rm in}| T_{t}{}^t| 0, {\rm in}\rangle \sim \frac{|\zeta (r_0)|^2}{(1-2M/r_0-\Lambda r_0^2/3)r_0^2}  \int_{0}^{\infty}
\frac{d\omega\, \omega\, f_{\omega} } {e^{\frac{2\pi \omega}{\kappa_H }} -1}   $$
Let $L_0$ be the cut-off in terms of the proper radius,
$$L_0= \int_{r_0}^{r_C} \frac{dr}{(1-2M/r-\Lambda r^2/3)^{1/2}} $$
 The smallest value of $L_0$ is expected to be the Planck length, $L_P$~\cite{Kolekar:2010py}. Writing $ (1-2M/r_0-\Lambda r_0^2/3) \approx \kappa_C (r_C-r)$ and recalling $|\zeta| \sim {\cal O}(1) $ near the horizon, we have 
 $$8 \pi G \,\langle 0, {\rm in}| T_{t}{}^t| 0, {\rm in}\rangle \sim \left( \frac{\kappa_H L_P }{\kappa_C L_0 r_0} \right)^2 $$
 Note that due to the appearance of the two surface gravities, the above term  is qualitatively different from a single horizon spacetime. Since we are ignoring backreaction of the field, we must have the upper bound for the 
sake of consistency,
$$ \left( \frac{\kappa_H L_P }{\kappa_C L_0 r_0} \right)^2 \lesssim \Lambda$$
Note that $\Lambda^{-1/2}$ sets one characteristic length scale of the theory whereas the other is set by $M$. But since we must have $M\sqrt{\Lambda} \leq 1$ in order to have two horizons, the left hand side of the above equation is automatically less than $M^{-2}$. Taking further $r_C \sim {\cal O}(\Lambda^{-1/2})$, we have at the leading order,
\begin{eqnarray}
L_0 \gtrsim \frac{\kappa_H}{\kappa_C} L_P
\label{estimate}
\end{eqnarray}
In the analogous expression for particles created by the cosmological horizon, the ratio $\kappa_H/\kappa_C$ is absent. Now for de Sitter black holes with comparable horizon sizes we may take $\kappa_H/\kappa_C \sim {\cal O}(1)$, giving $L_0 \gtrsim L_P$. Whereas for a few Solar mass black hole in our current universe we have $L_0 \gtrsim 10^{-5}{\rm m}$. As long as we are well above these bounds, our chosen `out' vacuum will not produce any considerable backreaction effects and we may safely use our `out' vacuum {\it till } $r_0$. Similar conclusion holds for the other components of $T_{\mu}{}^{\nu}$.

This completes our discussions on the Schwarzschild-de Sitter spacetime.
 Below we shall briefly discuss how these results generalise to the stationary axisymmetric Kerr-de Sitter spacetime.
\section{The case of the Kerr-de Sitter spacetime}\label{kds}
The Kerr-de Sitter metric in the Boyer-Lindquist coordinates reads,
\begin{eqnarray}
ds^2=-\frac{\Delta_r-a^2\sin^2\theta \Delta_{\theta}}{\rho^2}dt^2 -\frac{2a \sin^2 \theta }{\rho^2 \Xi} \left( (r^2+a^2 )\Delta_{\theta}-\Delta_r\right)dt d\phi \nonumber\\ + \frac{\sin^2 \theta }{\rho^2 \Xi^2} \left( (r^2 +a^2)^2 \Delta_{\theta}-\Delta_r a^2 \sin^2\theta\right)d\phi^2 + \frac{\rho^2}{\Delta_r}dr^2 + \frac{\rho^2}{\Delta_{\theta}}d\theta^2
\label{sup1}
\end{eqnarray}
where,
\begin{eqnarray*}
\Delta_r = (r^2 +a^2) \left(1-\frac{\Lambda r^2}{3}\right) -2Mr, \quad \Delta_{\theta} =1 + \frac{\Lambda a^2 \cos^2\theta}{3},  \quad \Xi = 1+ \frac{\Lambda a^2}{3},  \quad \rho^2 =r^2 +a^2 \cos^2 \theta 
\end{eqnarray*}
Setting $a=0$ above recovers the Schwarzschild-de Sitter spacetime whereas setting further $M=0$ recovers the de Sitter spacetime written in the static patch. Setting $M=0$ alone results in a line element diffeomorphic to the de Sitter, e.g.~\cite{Akcay:2010vt} and references therein.  The cosmological and the black hole event horizons as earlier are respectively given by the largest ($r_C$) and the next to the largest ($r_H$) roots of $\Delta_r=0$, whereas the smallest positive root, $r=r_-$ of $\Delta_r=0$ corresponds to the inner or the Cauchy horizon. There is an unphysical  negative root $r_u$ as well, $r_u=-(r_H+r_C+r_-)$. The surface gravities of the two horizons are respectively given by,
$$\kappa_{H,C}= \frac{\Delta_r'}{2(r^2+a^2)}\bigg\vert_{r=r_H,\, r_C}$$
Finally, the horizon Killing fields are given by $\chi_{H,C}=\partial_t+\Omega_{H,C}\,\partial_{\phi}$, with 
$$\Omega_{H,C}= \frac{a\, \Xi}{r_{H,C}^2+a^2}$$
being the angular speeds on the two horizons.\\

Defining the conserved quantities, $E=-g_{ab}(\partial_t)^a\,u^b$ and $L=g_{ab}(\partial_{\phi})^a u^b$ as earlier, we obtain the following equations for a null geodesic, e.g.~\cite{Bhattacharya:2017scw} and references therein,
\begin{eqnarray}
\frac{dt}{d\lambda}&=&\frac{\Xi}{\Delta_r \Delta_{\theta} \rho^2} \left[a\Delta_r\left(L-\frac{Ea\sin^2\theta }{\Xi} \right)+\frac{\Delta_{\theta} E (r^2+a^2)^2}{\Xi} \left(1 - \frac{aL\Xi}{E(r^2+a^2)} \right) \right]     \nonumber \\ 
\frac{d\phi}{d\lambda}&=&\frac{\Xi^2}{\Delta_r \Delta_{\theta }\rho^2 \sin^2\theta } \left[\Delta_r\left(L-\frac{Ea\sin^2\theta}{\Xi} \right) -La^2\sin^2\theta \Delta_{\theta}\left(1-\frac{E(r^2+a^2)}{a\Xi L} \right) \right]\nonumber \\
\rho^4\left(\frac{dr}{d\lambda}\right)^2 &=&   (r^2+a^2)^2\left(E-\frac{a\Xi L}{r^2+a^2}\right)^2-\Delta_r K_C,    \qquad \rho^4\left(\frac{d\theta}{d\lambda}\right)^2 = -\frac{1}{\sin^2\theta}\left(Ea\sin^2\theta - \Xi L\right) +\Delta_{\theta} K_C   \nonumber \\
\label{sup1'}
\end{eqnarray}
where $K_C$ is the Carter constant.  Unlike the static spacetime we cannot set $\phi = {\rm const.}$ here due to the frame dragging effect. Accordingly we shall study null geodesic moving along a constant $\theta$, say $\theta_0$ e.g.~\cite{Parker}. In that case we must have $K_C=0$ and in addition from the last of the above equations, $L=aE\sin^2 \theta_0/\Xi$. This simplifies the above equations to
\begin{eqnarray}
\frac{dt}{d\lambda}= \frac{E(r^2+a^2)}{\Delta_r}, \qquad \frac{d\phi}{d\lambda}= \frac{\Xi a E}{\Delta_r}, \qquad \frac{dr}{d\lambda} = \pm E
\label{sup1''}
\end{eqnarray}
where the $\pm$ sign in the last equation indicate outgoing and incoming geodesic respectively. The second of the above equations shows that the angular speed with respect to the affine parameter diverges on both the horizons. However in the locally non-rotating frames, $\phi_{H, C}= \phi -\Omega_{H,C}\, t $, it is easy to see that $d\phi_{H, C}/d\lambda=0$. In other words, these frames represent local observes rotating with the horizon, so that with respect to them the angular speed of the geodesic becomes vanishing.

If we now define the tortoise coordinate as, 
$$r_{\star}=\int dr \frac{(r^2+a^2)} {\Delta_r}$$
it is easy to see from the first and third of \ref{sup1''} that $u=t-r_{\star}$ and $v= t+r_{\star}$ are constants respectively 
along the outgoing and the incoming geodesics. Thus the  properties of the outgoing and incoming null geodesics for \ref{sup1} are essentially qualitatively the same as that of the Schwarzschild-de Sitter spacetime, \ref{ds11}. Accordingly, the expressions for the Kruskal coordinates are formally similar to \ref{ds15} as well.\\

By taking now the ansatz for the positive energy modes as,
$$\frac{e^{i(-\omega t+ m\phi)} S_{lm}(\theta) R_{\omega lm}(r)}{\sqrt{r^2+a^2}} $$
for the basis modes of the massless scalar field, one finds for the radial modes,
\begin{eqnarray}
\frac{d^2 R_{\omega lm}}{dr_{\star}^2} +\left(\omega - \frac{\Xi a m}{r^2+a^2} \right)^2 R_{\omega lm} +\frac{\Delta_r}{(r^2+a^2)^2} \left(-\frac{\Delta_r (2r^2-a^2)}{(r^2+a^2)^2} +\frac{r\Delta_r'}{r^2+a^2}  - \lambda_{lm} \right)R_{\omega lm}=0
\label{sup2}
\end{eqnarray}
where $S_{lm}(\theta)$'s  are the (orthonormal) spheroidal harmonics and $\lambda_{lm}$'s are corresponding eigenvalues~\cite{DeWitt}. In the near (future) black hole horizon limit, we have solutions, $R\sim e^{\pm i (\omega-m\Omega_H) r_{\star}}$. Defining the aforementioned azimuthal angular coordinate locally non-rotating on the horizon  $\phi_H:= \phi-\Omega_H t$, the  basis modes near the black hole horizon respectively take the form
\begin{eqnarray}
\frac{e^{-i(\omega - m\Omega_H)u} e^{im \phi_H}S_{lm}(\theta) }{\sqrt{r_H^2+a^2}}, \quad \frac{e^{-i(\omega - m\Omega_H)v} e^{i m \phi_H}S_{lm}(\theta) }{\sqrt{r_H^2+a^2}} 
\label{sup3}
\end{eqnarray}
If we restrict ourselves to  $\omega\geq m \Omega_H$ i.e., if we discard any superradiant
scattering~\cite{Wald:1984rg}, the above modes indeed become  positive frequency solutions. \\

In order to have the `in' mode functions on ${\cal C^-}$, we use a locally non-rotating coordinate system there, via $\phi=\phi_C+\Omega_C\, t$, which transforms \ref{sup1} as $r\to r_C$ as,
$$ds^2\approx -\frac{\Delta_r \,\rho^2}{(r^2+a^2)^2}dt^2 + \frac{\rho^2}{\Delta_r}dr^2+\frac{\rho^2}{\Delta_{\theta}}d\theta^2 +\frac{\sin^2\theta (r^2+a^2)^2 \Delta_{\theta}}{\rho^2 \Xi^2} d\phi_C^2$$
We expand the wave equation in the above background near the cosmological event horizon to obtain
\begin{eqnarray}
\left(-\frac{\partial^2 }{\partial t^2}+\frac{\partial^2 }{\partial r_{\star}^2}\right)p_{\omega \lambda m} +\frac{\Delta_r\,r^2}{(r^2+a^2)^{3}}\left(\frac{3\Delta_r}{r^2+a^2}-\frac{\Delta_r}{r^2}  -\frac{r \Delta_r'}{r^2} +\lambda_{lm} \right)p_{\omega \lambda m}=0
\label{sup4}
\end{eqnarray}
We may rewrite this equation using the Kruskal coordinates on the cosmological horizon, which are as we argued, formally similar to~\ref{ds15}. Accordingly, under the geometric optics approximation we obtain the expansion formally similar to~\ref{ds24} for the `in' modes (with $Y_{lm}$ replaced with $S_{lm}$ and $1/r_C$ replaced with $(r_C^2+a^2)^{-1/2}$) on ${\cal C^-}$. On the other hand, since the effective potential vanishes on the horizons, as earlier we find the `out' mode expansion formally similar to  \ref{ds25'}. Putting these all in together and using \ref{sup3} we obtain the analogue of \ref{ds26}
\begin{eqnarray}
 p^{(1)}_{\omega l m} &=& {\frac{e^{-i \omega v +i m \phi_C}}{\sqrt{2\pi} \sqrt{2 \omega}}}\,\frac{ S_{lm} (\theta)}{(r_C^2+a^2)^{1/2}}={\frac{e^{ \frac{i\omega}{\kappa_C} \ln (-\kappa_C V_C) +i m \phi_C}}{\sqrt{2\pi}\sqrt{2 \omega}}}\,\frac{S_{lm} (\theta)}{(r_C^2+a^2)^{1/2}}~~({\rm modes~that~did~not~enter~the~collapsing~body} )\nonumber \\
 p^{(2)}_{\omega l m} &=& {\frac{e^{ \frac{i(\omega-m\Omega_H)}{\kappa_H}\ln (C_0 (V_C-V_0)) +im \phi_C }}{\sqrt{2\pi} \sqrt{2 \omega}}}\,\frac{S_{lm} (\theta)}{(r_C^2+a^2)^{1/2}}~~({\rm entered~and~got~scattered~just~prior~to}~{\cal H^+}~{\rm formed})
\label{ds26kerr}
\end{eqnarray}

The rest of the computation follows in an exactly similar manner as the static spacetime described in \ref{bogoliu}. The near-${\cal C^+}$ cut-off $L_0$ introduced in \ref{bogoliu} reads here as
$$L_0= \int_{r_0}^{r_C} \frac{(r^2+a^2)^{1/2} \,dr}{\sqrt{\Delta_r}} $$
The particle creation rate  by the cosmological event horizon turns out to be the same as~\ref{ds31} whereas   for the black hole we obtain 
\begin{eqnarray}
n^{H}_{\omega l m}= \frac{1}{2\pi}\frac{f_{\omega lm }}{e^{\frac{2\pi (\omega-m \Omega_H)}{\kappa_H}}-1} \qquad (\omega-m \Omega_H>0)
\label{ds30sa}
\end{eqnarray}

The above results indicate the existence of thermal fluxes as earlier. However, any explicit computation of the expectation values of the energy-momentum tensor will be much harder compared to that of \ref{ds1}, for the angular eigenvalues ($\lambda_{l m}$ in \ref{sup2}) will now be highly non-trivial due to the absence of the spherical symmetry e.g.,~\cite{Casals:2012es, daCunha:2015fna} and references therein.  To the best of our knowledge, there has been no systematic study yet of the vacuum expectation values of the energy-momentum tensor for various spin fields in the Kerr-de Sitter geometry.    We hope to address this issue in  full detail in our future  works. However in any case it is evident that  the thermal characteristics of the outgoing spectrum will remain indeed intact.

\section{Discussions}\label{disc}
In this work we considered particle creation in a quasistationary gravitational collapse occurring within the region enclosed by the past and future segments (${\cal C^-}$ and ${\cal C^+}$) of the cosmological event horizon  in a $(3+1)$-dimensional eternal de Sitter universe, \ref{fig2}.  We considered the dynamics of a massless minimally coupled scalar field obeying the null geodesic approximation.
The only reasonable `in' vacuum was the one defined with respect to the positive frequency regular Kruskal $V_C$ modes localised on ${\cal C^-}$ at the onset of the collapse, \ref{ds24}. The usual $(u,v)$ modes were used to describe the collapse dynamics and the `out' vacuum was defined with respect to the positive frequency mode $\sim e^{-i\omega u}$ at some radial point `close' to ${\cal C^+}$. With these choices of vacua, we computed  after ray tracing to ${\cal C^-}$ the Bogoliubov coefficients and   the two temperature spectra of created particles in \ref{bogoliu}.  
 Since the coordinate $u$ used to define the `out' vacuum is not well defined on ${\cal C^+}$, we have made a heuristic estimate by considering backreaction at the end of~\ref{bogoliu}, on how close we might be to it while defining that vacuum.  We  further have provided arguments in the favour of uniqueness of the `out' vacuum we have chosen, in order the obtain particle creation by both horizons.
 
 Note that there will also be an influx of particles at late times toward ${\cal H^+}$, \ref{ds25'}. The spectrum will depend upon the precise form of the $q_{\omega l m}$ modes we choose. Also, as long as we are not taking the Nariai limit $r_H\to r_C$, our results hold good irrespective of the relative sizes of the two horizons.

The chief qualitative differences of this derivation with that of the eternal black holes~\cite{Traschen:1999zr, Choudhury:2004ph} seems to be the difference in the number of `in' vacuum. First, for such black holes we must have two Kruskal vacua associated with the two horizons which are the analogues of the `in' vacuum.   In our present case as we discussed earlier, there can be only one `in' vacuum and the two temperature spectra is derived with respect to that vacuum only.  

There exist several interesting and important directions that could be pursued further. First, it would be interesting to see if we can relax the eternal characteristics of the de Sitter horizon as well. This would probably correspond to formation of a black hole in a $\Lambda{\rm CDM}$ universe described by a McVittie metric, e.g.~\cite{Bhattacharya:2017bpl} and references therein (see also \cite{Saida:2007ru} and references therein for a discussion on particle creation in a Swiss-cheese universe). Second, as we have discussed at the end of \ref{kds}, the computation of the regularised vacuum expectation values of the energy-momentum tensors of various spin fields in the Kerr-de Sitter background seems to be very important as well. Finally, it would further be important to understand all these results in the Nariai limit important especially in the early universe scenario ~\cite{Bousso:1997wi, Chao:1997em,  Bousso:1999ms, Anninos:2010gh}. We hope to address these issues in our future works.

\vskip 1cm


\section*{Acknowledgements}
I would like 
 thank Moutushi for assisting me to draw the diagramms. This research is partially supported by the ISIRD grant 9-289/2017/IITRPR/704.


\end{document}